\documentclass[twocolumn,showpacs,preprintnumbers,amsmath,amssymb,prb]{revtex4}

\usepackage{graphicx}
\usepackage{dcolumn}
\usepackage{bm}
\usepackage{color}

\begin{document}

\preprint{}

\title{Elastic Instabilities within Antiferromagnetically Ordered Phase in the Orbitally-Frustrated Spinel GeCo$_2$O$_4$}

\author{Tadataka Watanabe$^1$}
\author{Shigeo Hara$^2$}
\author{Shin-Ichi Ikeda$^3$}
\author{Keisuke Tomiyasu$^4$}
\affiliation{$^1$Department of Physics, College of Science and Technology (CST), Nihon University, Chiyoda, Tokyo 101-8308, Japan}
\affiliation{$^2$Department of Physics, Chuo University, Bunkyo, Tokyo 101-8324, Japan}
\affiliation{$^3$Nanoelectronics Research Institute, National Institute of Advanced Industrial Science and Technology (AIST), Tsukuba, Ibaraki 305-8568, Japan}
\affiliation{$^4$Department of Physics, Tohoku University, Aoba, Sendai 980-8578, Japan}
\date{\today}

\begin{abstract}
Ultrasound velocity measurements of the orbitally frustrated spinel GeCo$_2$O$_4$ reveal colorful and unusual elastic anomalies within the antiferromagnetic phase. Temperature dependence of shear moduli exhibits a minimum within the antiferromagnetic phase, suggesting the coupling of shear acoustic phonons to molecular spin-orbit excitations. Magnetic-field dependence of elastic moduli exhibits diplike anomalies, being interpreted as magnetic-field-induced metamagnetic and structural transitions. These elastic anomalies suggest that the survival of geometrical frustration, and the interplay of spin, orbital, and lattice degrees of freedom evoke a rich set of phenomena in the antiferromagnetic phase.
\end{abstract}


\maketitle

The concept of geometrical frustration provides an intriguing playground for condensed matter physics. So far, the majority of studies have been devoted to the spin degrees of freedom \cite{Moessner}. Here, we would like to focus on the orbital sector as the leading role. An orbital system is inherently frustrated even on a simple square lattice: when orbitals (directions of the electron cloud) are arranged to gain bond energy for one direction, this configuration is not fully favorable for other bonds \cite{Khomskii, Tanaka, Nasu}. Furthermore, the orbital frustration is expected to be enhanced on a geometrically frustrated lattice, where the orbitals as well as spins cannot form antiferro-type order \cite{Khomskii,Buttgen}.

Cubic spinels $AB_2$O$_4$ with magnetic $B$ ions have attracted much interest in light of the geometrical frustration which is inherent in the $B$-site sublattice of corner sharing tetrahedra. The spinel cobaltite GeCo$_2$O$_4$ consists of magnetic Co$^{2+}$ (3$d^7$) on the octahedral $B$ sites and non-magnetic Ge$^{4+}$ on the tetrahedral $A$ sites. The positive Weiss temperature, $\Theta_W$ = +81.0 K, indicates the dominant contribution of ferromagnetic (FM) interactions \cite{Diaz1}. However, an antiferromagnetic (AF) long-range order appears below $T_N$ = 21.0 K, in coincidence with a cubic-to-tetragonal structural elongation with $c/a$ = 1.001 \cite{Hubsch, Diaz2, Hoshi}. The AF order is composed of a dominant component described by a trigonal propagation vector $\vec{Q}_{II} = (1/2, 1/2, 1/2)$ and an additional one described by a tetragonal $\vec{Q}_{I} = (1, 0, 0)/(0, 1, 0)$, where the magnitude of $\vec{Q}_I$ is much weaker than that of $\vec{Q}_{II}$ \cite{Hubsch, Diaz2, Tomiyasu2}.

GeCo$_2$O$_4$ is a promising candidate for the frustrated system with orbital degrees of freedom \cite{Watanabe, Tomiyasu1}. Ultrasound velocity measurements suggested that the structural transition at $T_N$ in GeCo$_2$O$_4$ is ascribed to the release of frustration rather than the Jahn-Teller effect \cite{Watanabe}. An octahedral-site Co$^{2+}$ is a well-known spin-orbit (SO)-coupling active ion with an unquenched orbital angular momentum $\vec{L}$. The electronic ground state is normally described by effective total angular momentum $J_{eff}=1/2$ doublet with $L=1$ based on the triply-degenerate $t_{2g}$ orbitals and $S=3/2$ \cite{Kanamori, Lines}. In GeCo$_2$O$_4$, the $J_{eff}=1/2$ ground doublet was confirmed by specific heat measurements and inelastic neutron scattering experiments \cite{Lashley, Tomiyasu1}.

An inelastic neutron scattering study discovered that a strong frustration effect (molecular spin excitations) still survives even below $T_N$ in the typical spin-frustrated spinel chromite MgCr$_2$O$_4$ \cite{Tomiyasu3}. Very recently, it was also reported that GeCo$_2$O$_4$ exhibits molecular SO excitations, not only above but also below $T_N$ \cite{Tomiyasu1}. Therefore, novel phenomena relevant to the geometrical frustration are expected to be hidden in the AF state in GeCo$_2$O$_4$.

In this paper, we present ultrasound velocity measurements in GeCo$_2$O$_4$. The modified sound dispersions by the magneto-elastic coupling allow one to extract detailed information about the interplay of lattice, spin, and orbital degrees of freedom. We find colorful and unusual elastic anomalies in GeCo$_2$O$_4$ in the AF state.

Ultrasound velocities were measured in single crystals of GeCo$_2$O$_4$ prepared by the floating zone method \cite{Hara}, where the phase comparison technique was utilized with longitudinal and transverse sound waves at a frequency of 30 MHz. The ultrasound waves were generated and detected by LiNbO$_3$ transducers glued on the parallel mirror surfaces of the crystal. We measured sound velocities in all the symmetrically independent elastic modes in the cubic crystal: compression modulus $C_{11}$ ($A_{1g}$ symmetry), tetragonal shear modulus $\frac{(C_{11}-C_{12})}{2}\equiv C_t$ ($E_g$ symmetry), and trigonal shear modulus $C_{44}$ ($T_{2g}$ symmetry).

Figures 1(a) and (b) depict the transverse sound velocity $v_T$ in $C_{44}$ and $C_t$ with magnetic field $H \|$[110] as functions of temperature ($T$), respectively. The theoretical extension of the experimental data in the paramagnetic (PM) region of 50 K $<T <$150 K down to $T\rightarrow0$, the background $C_{ij}^0$, is also shown as the dotted curve \cite{Varshni}. As reported in our previous study \cite{Watanabe}, $C_{44}(T)$ and $C_t(T)$ with $H$ = 0 exhibit, in addition to a diplike anomaly at $T_N$ = 20.6 K, nonmonotonic $T$ dependence in $T<T_N$ with a characteristic minimum at $\sim$6 K \cite{Watanabe}. Furthermore, as a newly discovered feature, $C_t(T)$ at 2 T exhibits the steep softening below $\sim$6 K with decreasing $T$.

We explain that the minimum in $C_{44}(T)$ and $C_t(T)$ with $H$ = 0, and the softening in $C_t(T)$ at 2 T arise from magnetic origins, not from domain-wall effect. In the magnetically ordered state, magnetostriction can give rise to the domain-wall stress which is detected as loss of elasticity \cite{Domain}. This stress is generally removed by applying $H$, which is observed as recovery of elasticity with $H$. For GeCo$_2$O$_4$, the domain-wall stress effect on the volume-strain mode $C_{11}$ should be stronger than the volume-conserving modes $C_{44}$ and $C_t$. Fig. 1(c) depicts $T$ dependence of the longitudinal sound velocity $v_L$ in $C_{11}$. Below $T_N$, $C_{11}(T)$ with $H$ = 0 loses the elasticity compared to 3 T. Thus $C_{11}(T)$ with $H$ = 0 should be affected by the domain-wall stress. However, $C_{11}(T)$ with $H$ = 0 is almost constant below $T_N$, indicating the $T$ independence of the domain-wall stress effect.

\begin{figure}[t]
\begin{center}
\includegraphics[scale=0.33]{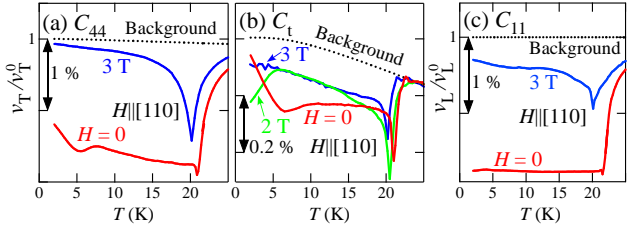}
\caption{\label{fig:Fig2} (Color online). $T$ dependence of $v_T$ and $v_L$ with $H \|$[110]. (a) $C_{44}$, (b) $\frac{(C_{11}-C_{12})}{2}\equiv C_t$, and (c) $C_{11}$. The dotted curve indicates the background $C_{ij}^0$ in each modulus \cite{Varshni}. The absolute value of the sound velocity at 2 K is $v_T^0$ = 4030 m/s for $C_{44}$, $v_T^0$ = 3300 m/s for $C_t$, and $v_L^0$ = 6940 m/s for $C_{11}$, respectively.}
\end{center}
\end{figure}

Figures 2(a) and (b) depict the relative change of $v_T$ in $C_{44}$ and $C_t$ as functions of $H \|$[110], respectively. $C_{44}(H)$ and $C_t(H)$ exhibit several distinct anomalies within the AF phase at 2.0 K and 17.5 K, in contrast to the absence in the PM phase at 30 K. At 17.5 K, a diplike anomaly is seen in $C_{44}(H)$ and $C_t(H)$ at $\mu_0 H_N$ = 6.6 T as marked by the open arrows, which coincides with the AF-to-PM transition observed in the magnetization as well as the specific heat measurements \cite{Hoshi, Hoshi2}. Below $H_N$ at 17.5 K, $C_{44}(H)$ exhibits steep hardening with increasing $H$ in $\sim$1 T$<\mu_0 H<\sim$3 T, in contrast to the weak $H$ dependence in $C_t(H)$. At 2.0 K, $H_N$ shifts to over 7 T, and further anomalies evolve as marked by the filled/open circles.

Although the isothermal magnetization measurements at 1.5 K with $H||$[110] reported a step-like anomaly at $\sim$4 T suggesting the $H$-induced transition \cite{Hoshi}, no $C_{ij}(H)$ at 2 K in Figs. 2(a)-(c) exhibits anomaly at $\sim$4 T. Very recently, neutron scattering experiments under $H$ identified this $\sim$4 T anomaly as a spin reorientation transition in $\vec{Q}_{II}$ \cite{Matsuda}. The absence of anomaly in $C_{ij}(H)$ suggests that, in contrast to the magnetic susceptibility $-\partial^2 F/\partial H^2$, the strain susceptibility $\partial^2 F/\partial \epsilon^2$ is insensitive to the $H$-induced transition at $\sim$4 T in $\vec{Q}_{II}$. In fact, the trigonal $\vec{Q}_{II}$ is inconsistent with the tetragonal lattice distortion in symmetry, suggesting weakness of the coupling between $\vec{Q}_{II}$ and lattice.

\begin{figure}[b]
\begin{center}
\includegraphics[scale=0.33]{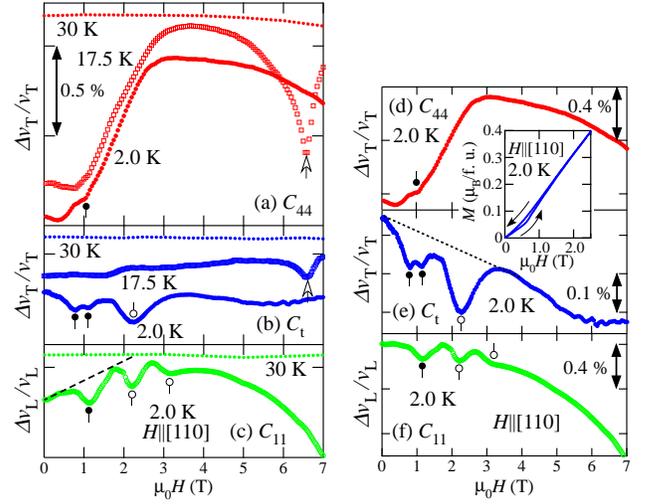}
\caption{\label{fig:Fig2} (Color online). (a)-(c) The relative shift of $v_T$ and $v_L$ as functions of $H \|$[110]. (a) $C_{44}$, (b) $C_t$, and (c) $C_{11}$. The curves are shifted for clarity. The open arrows in (a) and (b) indicate the AF-to-PM transition at 17.5 K. The dashed line in (c) is a linear fit to $C_{11}(H)$ at 2.0 K in 0 $<\mu_0 H<$ 0.5 T. (d)-(f) The data at 2.0 K in (a)-(c) after the subtraction of the domain-wall stress effect. The dotted line in (e) is a guide to the eye indicating the softening in $C_t(H)$. Inset shows the magnetization curve at 2.0 K. The filled/open circles in (a)-(f) indicate the elastic anomalies with/without the magnetic hysteresis.}
\end{center}
\end{figure}

Assuming that the domain-wall stress effect in $C_{11}(H)$ at 2.0 K is expressed by the linear fit in $0<\mu_0 H<0.5$ T as indicated by the dashed line in Fig. 2(c), Figs. 2(d)-(f) depict $C_{44}(H)$, $C_t(H)$, and $C_{11}(H)$ at 2.0 K with the subtraction of the domain-wall stress effect, respectively. $C_{44}(H)$ exhibits the hardening with increasing $H$ in $\sim$0.5 T$<\mu_0 H<\sim $3 T. Referring to $C_{44}(T)$ in Fig. 1(a), this hardening in $C_{44}(H)$ corresponds to the recovery towards the background elasticity with $H$, $C_{44}^0-C_{44}\equiv\Delta C_{44}\rightarrow$ 0. Thus the most probable origin for the hardening in $C_{44}(H)$ is the suppression of the magneto-elastic coupling with increasing $H$. In contrast, $C_t(H)$ exhibits the softening with increasing $H$ in 0 $<\mu_0 H< \sim $6 T as indicated by the dotted line in Fig. 2(e), which corresponds to the increase of $C_t^0-C_t\equiv\Delta C_t$ with increasing $H$. Thus the softening in $C_t(H)$ indicates that, in contrast to $C_{44}(H)$, the magneto-elastic coupling is enhanced with increasing $H$. Such elastic-mode-dependent $H$ variations of the elastic constants, the hardening in $C_{44}(H)$ and the softening in $C_t(H)$, are observed most likely as the results of the coupling of the sound waves to the spin-wave excitations \cite{Melcher}. At $H$ = 0, referring to Figs. 1(a) and (b), $\Delta C_{44}/C_{44}$ is much larger than $\Delta C_t/C_t$, indicating that the spin-wave excitations affect $C_{44}$ much more strongly than $C_{t}$. The downturn softening in $C_{44}(H)$ and $C_{11}(H)$ in $\sim$3 T$<\mu_0 H<$7 T would be the precursor to the AF-to-PM transition at the higher $H$.

In Figs. 2(d)-(f), another salient feature is the diplike anomalies marked by the filled/open circles, indicative of the $H$-induced transitions. These anomalies evolve in the lower $T$ in the AF state, and are elastic-mode selective with the occurrence at $\sim$1 T in all the elastic moduli, at $\sim$2 T in $C_t$ and $C_{11}$, and at $\sim$3 T in $C_{11}$. Here, we regard the shoulder structure in $C_{44}(H)$ at $\sim$1 T as a diplike anomaly because this structure should be formed by the superposition of the hardening in $\sim$0.5 T$<\mu_0 H<\sim $3 T on the diplike feature. The diplike anomaly in $C_{ij}(H)$ within the magnetically ordered state is normally observed at the metamagnetic transition \cite{Luthi}. The inset to Figs. 2(d)-(f) shows the magnetization curve at 2.0 K with $H \|$[110], where the change of slope and the hysteretic feature are seen in $0<\mu_0 H<\sim$1 T. Thus the elastic anomalies at $\sim$1 T in Figs. 2(d)-(f) marked by the filled circles should be attributed to the metamagnetic transition. In contrast, the elastic anomalies at $\sim$2 T and $\sim$3 T in Figs. 2(e) and (f) marked by the open circles are attributed to other $H$-induced transitions without anomaly in the magnetization curve.

From now on, let us discuss in detail the origins of the two types of unusual elastic anomalies observed within the AF phase: a characteristic minimum in $C_{44}(T)$ and $C_t(T)$ with $H$ = 0, and diplike anomalies in $C_{44}(H)$, $C_t(H)$, and $C_{11}(H)$. First, we discuss the origin of a minimum at $\sim$6 K in $C_{44}(T)$ and $C_t(T)$ with $H$ = 0 shown in Figs. 1(a) and (b). According to the specific heat measurements in GeCo$_2$O$_4$, there is no additional phase transition within the AF phase with $H$ = 0 \cite{Lashley}. Thus the minimum in $C_{44}(T)$ and $C_t(T)$ with $H$ = 0 should originate from the coupling of the sound waves (acoustic phonons) to the magnetic excitations, not to the static order. As the possible magnetic excitations, we can first consider the spin-wave excitations. In GeCo$_2$O$_4$, as described above in conjunction with Figs. 2(d) and (e), the coupling of the sound wave to the spin-wave excitations should be suppressed in $C_{44}$ but $enhanced$ in $C_t$ with increasing $H$. However, as exemplified by the data at 3 T in Figs. 1(a) and (b), the minimum disappears by the application of $H>$ 2 T in not only $C_{44}(T)$ $but$ $also$ $C_t(T)$ \cite{Watanabe}. Thus we rule out the spin-wave excitations as the origin of the minimum in $C_{44}(T)$ and $C_t(T)$ with $H$ = 0. We now note that the inelastic neutron scattering study in GeCo$_2$O$_4$ discovered the appearance of an energy gap $\Delta\sim$ 3 meV as the low-energy magnetic excitations in the AF state \cite{Tomiyasu1,Lashley}. And, interestingly, this 3 meV-gapped mode was characterized as the molecular SO excitations which coexist with the AF order \cite{Tomiyasu1}. Therefore, {\it the minimum in $C_{44}(T)$ and $C_t(T)$ with $H$ = 0 at $T<T_N$ should stem from the molecular SO excitations}. Noteworthily, this conclusion means that the elastic anomalies due to the molecular magnetic excitations (frustration effect) have been observed, as far as we know, for the first time.

\begin{figure}[b]
\begin{center}
\includegraphics[scale=0.33]{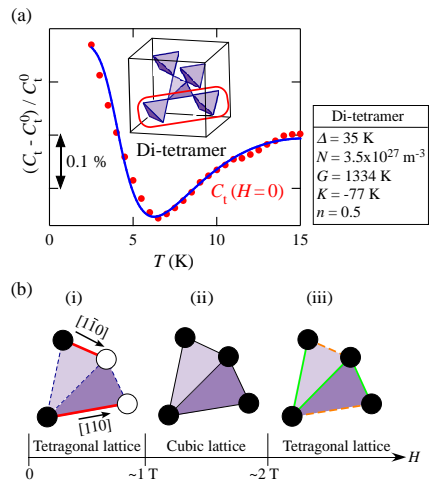}
\caption{\label{fig:Fig4} (Color online). (a) $T$ dependence of $C_t$ with $H$ = 0 from Fig. 1(b) (symbols). The solid curve is a fit to Eq. (1) in assumption of the molecular SO excitations $\Delta$ = 35 K. Inset picture shows a di-tetramer in the $B$-site Co$^{2+}$ sublattice. The fit parameters are summarized in the right table. (b) $H$-induced metamagnetic and structural transitions in the $\vec{Q}_I$ order schematically depicted in a Co$^{2+}$ tetrahedron. The filled and open circles depict up and down spins.}
\end{center}
\end{figure}

Here, we would like to give a quantitative discussion of the molecular SO excitations $\Delta$ using $C_t(T)$ with $H$ = 0 in Fig. 1(b), shown again in Fig. 3(a) (symbols). The inelastic neutron scattering study in GeCo$_2$O$_4$ suggested that, in the ground state, the nonmagnetic singlet pairs of the FM Co$^{2+}$ tetrahedrons, "di-tetramers" shown in the inset to Fig. 3(a), coexist with the AF order \cite{Tomiyasu1}. Here, $\Delta\sim$ 3 meV ($\sim$35 K) corresponds to the singlet$\rightarrow$triplet excitation energy of the di-tetramers \cite{Tomiyasu1}. A characteristic minimum in $C_t(T)$ with $H=0$ due to such short-range magnetic excitations is analogous to the one in the low-dimensional spin dimer systems like SrCu$_2$(BO$_3$)$_2$ \cite{Wolf}. The contribution of the di-tetramers to the elastic constant is written as,
\begin{equation}
C_t=C_t^0-G^2N\frac{\chi_s}{(1-K\chi_s)},
\label{eq:ES1}
\end{equation}
with $N$ the density of di-tetramers, $G=|\partial \Delta/\partial \epsilon|$ the coupling constant to a single di-tetramer which measures the strain dependence of the singlet-triplet gap, $K$ the inter-di-tetramer interaction, and $\chi_s=G^{-2}(\partial^2 F/\partial \epsilon^2)$ the strain susceptibility of a single di-tetramer. Since $\Delta$ is expected to evolve in the lower $T$ from the ungapped state at $T_N$ \cite{Hoshi,Tomiyasu1,Lashley}, we assume the $T$-dependent gap, $\Delta(T) = \Delta (1-(T/T_N)^{n})$. Here, the value of $\Delta$ = 35 K is fixed as the first approximation. In Fig. 3(a), it is evident that a fit to Eq. (1) shown as the solid curve excellently reproduces a minimum at $\sim$6 K in $C_t(T)$ supporting the presence of the molecular SO excitations. The large fitted value of $G$ = 1334 K indicates that the dimerized FM tetrahedrons (di-tetramers) in the tetragonal AF phase strongly couple to the lattice deformation. On the other hand, the absence of anomaly in $C_t(T)$ in the PM phase \cite{Watanabe} indicates $G\simeq$ 0 for the gapless molecular state of the undimerized FM tetrahedrons in the cubic PM phase \cite{Tomiyasu1}. These features imply the correlation between the lattice structure and the SO molecular state.

Second, we discuss the origins of the diplike anomalies in $C_{44}(H)$, $C_t(H)$, and $C_{11}(H)$ at 2 K shown in Figs. 2(d)-(f). For a sound wave with polarization $u$ and propagation $k$, the magneto-elastic coupling acting on the exchange interactions is written as \cite{Luthi},
\begin{equation}
H_{exs} = \sum_{i}(\frac{dJ}{d\delta} \cdot u)(k \cdot \delta)(S_i \cdot S_j)e^{i(k \cdot R_i-\omega t)}.
\label{eq:ES2}
\end{equation}
Here $\delta = R_i-R_j$ is the distance between two magnetic ions. On the basis of this formula, the metamagnetic transition at $\sim$1 T observed in $C_{44}$ ($k||[001]$, $u||[110]$), $C_t$ ($k||[110]$, $u||[1\bar{1}0]$), and $C_{11}$ ($k||u||[100]$) should be dominated by the exchange interactions in $<$110$>$ direction because the exchange interactions in $<$100$>$ and $<$111$>$ directions become inactive in $C_{44}$ and $C_t$, respectively. Thus the metamagnetic transition at $\sim$1 T should occur in the $\vec{Q}_{I}$ order which is composed of the antiferromagnetic spin chains along [110] and [1$\bar{1}$0] directions as shown in Fig. 3(b)(i) \cite{Tomiyasu2}. Note that this metamagnetic transition leads to the disappearance of the $\vec{Q}_{I}$ order, as shown in Fig. 3(b)(ii). Therefore, it is expected that the tetragonal-to-cubic structural transition coincides with the metamagnetic transition at $\sim$1 T, as shown in Figs. 3(b)(i) and (ii).

The $H$-induced transition at $\sim$2 T is characterized by the anomaly in $C_t(H)$ as shown in Fig. 2(e) (open circle), the evolution of which corresponds to the softening in $C_t(T)$ at 2 T below $\sim$6 K, as shown in Fig. 1(b). The absence of anomaly at $\sim$2 T in the isothermal magnetization, as shown in the inset to Figs. 2(d)-(f), rules out the possible metamagnetic transition. Thus the anomaly in the tetragonal $C_t$ at $\sim$2 T strongly suggests the occurrence of the cubic-to-tetragonal structural transition, as shown in Figs. 3(b)(ii) and (iii). The anomaly at $\sim$3 T only in $C_{11}(H)$ ($A_{1g}$ symmetry) suggests, with the absence of anomaly in the isothermal magnetization \cite{Hoshi}, the occurrence of the isostructural non-metamagnetic transition. The nature of this transition is uncovered at this stage. Taking into account the SO-coupling active Co$^{2+}$, it is expected that the $H$-induced structural transition at $\sim$2 T coincides with the orbital state transition driven by $H$ and SO coupling. Detailed studies by applying the microscopic probes such as NMR, neutron scattering, and resonant X-ray scattering would uncover the exotic orbital phenomena.

As mentioned above in conjunction with Figs. 1(a) and (b), the minimum in $C_t(T)$ and $C_{44}(T)$ with $H$ = 0 disappears in the data with $H>$ 2 T, indicating the decoupling of the shear acoustic phonons from the molecular SO excitations. This $H$-induced decoupling might be driven by the $H$-induced structural transition at $\sim$2 T. Here, the transformation or the disappearance of the SO molecules is expected to coincide with the $H$-induced structural transition.

In summary, ultrasound velocity measurements of GeCo$_2$O$_4$ reveal unusual elastic anomalies within the AF phase due to the molecular SO excitations and the $H$-induced metamagnetic and structural transitions. The present results suggest that the survival of geometrical frustration, and the interplay of spin, orbital, and lattice degrees of freedom evoke a rich set of phenomena in the AF phase which deserves further experimental and theoretical studies.

This work was partly supported by Grants-in-Aid for Young Scientists (B) (21740266) and Scientific Research on Priority Areas (22014001) from MEXT of Japan.


\begin{thebibliography}{prb}
\bibitem{Moessner} R. Moessner and A. Ramirez, Phys. Today {\bf 59}, 24 (2006).
\bibitem{Khomskii} D. I. Khomskii and M. V. Mostovoy, J. Phys. A: Math. Gen. {\bf 36}, 9197 (2003).
\bibitem{Tanaka} T. Tanaka and S. Ishihara, Phys. Rev. Lett. {\bf 98}, 256402 (2007).
\bibitem{Nasu} J. Nasu, A. Nagano, M. Naka, and S. Ishihara, Phys. Rev. B. {\bf 78}, 024416 (2008).
\bibitem{Buttgen} N. B$\ddot{u}$ttgen, A. Zymara, C. Kegler, V. Tsurkan, and A. Loidl, Phys. Rev. B {\bf 73}, 132409 (2006).
\bibitem{Diaz1} S. Diaz S. de Brion, M. Holzapfel, G. Chouteau, and P. Strobel, Physica B {\bf 346-347}, 146 (2004).
\bibitem{Hubsch} J. Hubsch and G. Gavoille, J. Magn. Magn. Mater. {\bf 66}, 17 (1987).
\bibitem{Diaz2} S. Diaz, S. de Brion, G. Chouteau, B. Cannals, V. Simonet, and P. Strobel, Phys. Rev. B {\bf 74}, 092404 (2006).
\bibitem{Hoshi} T. Hoshi, H. Aruga-Katori, M. Kosaka, and H. Takagi, J. Magn. Magn. Mater. {\bf 310}, e448 (2007).
\bibitem{Tomiyasu2} K. Tomiyasu, A. Tominaga, S. Hara, H. Sato, T. Watanabe, S. I. Ikeda, H. Hiraka, K. Iwasa, and K. Yamada, J. Phys.: Conf. Ser. (to be published).
\bibitem{Watanabe} T. Watanabe, S. Hara, and S. I. Ikeda, Phys. Rev. B {\bf 78}, 094420 (2008).
\bibitem{Tomiyasu1} K. Tomiyasu, M. K. Crawford, D. T. Adroja, P. Manuel, A. Tominaga, S. Hara, H. Sato, T. Watanabe, S. I. Ikeda, J. W. Lynn, K. Iwasa, and K. Yamada, arXiv:1102.4395 (unpublished).
\bibitem{Kanamori} J. Kanamori, Prog. Theor. Phys. {\bf 17}, 177 (1957).
\bibitem{Lines} M. E. Lines, Phys. Rev. {\bf 131}, 546 (1963).
\bibitem{Lashley} J. C. Lashley, R. Stevens, M. K. Crawford, J. Boerio-Goates, B. F. Woodfield, Y. Qin, J. W. Lynn, P. A. Goddard, and R. A. Fisher, Phys. Rev. B {\bf 78}, 104406 (2008).
\bibitem{Tomiyasu3} K. Tomiyasu, H. Suzuki, M. Toki, S. Itoh, M. Matsuura, N. Aso, and K. Yamada, Phys. Rev. Lett. {\bf 101}, 177401 (2008).
\bibitem{Hara} S. Hara, Y. Yoshida, S. I. Ikeda, N. Shirakawa, M. K. Crawford, K. Takase, Y. Takano, and K. Sekizawa, J. Crystal Growth {\bf 283}, 185 (2005).
\bibitem{Varshni} V. P. Varshni, Phys. Rev. B {\bf 2}, 3952 (1970).
\bibitem{Domain} R. M. Bozorth, {\it Ferromagnetism} (Van Nostrand, Princeton, 1951).
\bibitem{Hoshi2} T. Hoshi, H. Aruga-Katori, M. Kosaka, and H. Takagi, unpublished.
\bibitem{Matsuda} M. Matsuda, T. Hoshi, H. Aruga-Katori, M. Kosaka, and H. Takagi, J. Phys. Soc. Jpn. {\bf 80}, 034708 (2011).
\bibitem{Melcher} R. L. Melcher and D. I. Bolef, Phys. Rev. {\bf 186}, 491 (1969).
\bibitem{Luthi} B. L$\ddot{u}$thi, {\it Physical Acoustics in the Solid State} (Springer, 2005).
\bibitem{Wolf} B. Wolf, S. Zherlitsyn, S. Schmidt, B. L$\ddot{u}$thi, H. Kageyama, and Y. Ueda, Phys. Rev. Lett. {\bf 86}, 4847 (2001).
\end{thebibliography}
\end{document}